# Quantum Secret Sharing with error correction

[1]Aziz Mouzali, [2]Fatiha Merazka, [3]Damian Markham

[1]Department of physics, ENP, Algiers, Algeria.
[2]Electronic & Computer Science Faculty, USTHB, Algeria.
[3]Laboratoire Traitement et Communication de l'Information.
CNRS - Télécom ParisTech, France.

## Abstract

We investigate in this work a quantum error correction on a five-qubits graph state used for secret sharing through five noisy channels. We describe the procedure for the five, seven and nine qubits codes. It is known that the three codes always allow error recovery if only one among the sents qubits is disturbed in the transmitting channel. However, if two qubits and more are disturbed, then the correction will depend on the used code. We compare in this paper the three codes by computing the average fidelity between the sent secret and that measured by the receivers. We will treat the case where, at most, two qubits are affected in each one of five depolarizing channels.

Keywords: Quantum Correction, Graph State, Quantum Secret Sharing, Feynman Program.

PACS numbers : 03.67.Acn, 03.67.Hk, 03.67.Lx

## 1. Introduction

The graph state can be very useful for several quantum protocols as secret sharing, measurement-based computation, error correction, teleportation and quantum communications. Then, it would be in the future a good way to unify these topics in one formalism. For example, an output of quantum computation considered as secret can be included in a graph state, then protected by a quantum error correcting code and sent trough noisy channel to several receivers sharing this secret. The quantum secret sharing with graph state is very well described in [1], particularly the five qubits graph state. In this last case, only three receivers among five will access the secret, the two others being considered as eavesdroppers. In this work, we investigate the effects of the five, seven and nine qubits codes used to protect a five qubits graph state containing a secret and sent by a dealer to five players. We will compare the fidelity to determine the best code for a depolarizing channel where only one or two sent qubits are error affected. Some of the results have been obtained using a simulator called "Feynman Program", witch is a set of procedures supporting the definition and manipulation of an n-qubits system and the unitary gates acting on them. This program is described in details in [2][3][4][5] and obtainable from [6].

## 2. Quantum Secret Sharing

Quantum secret sharing (QSS) is a quantum cryptographic protocol wherein a dealer shares private or public quantum channels with each player, while the players share private quantum or classical channels between each other. The dealer prepares an encoded version of the secret using a qubits string which he transmits to n players, only a subset k of them can collaborates to reconstruct the secret. We call a (k,n) threshold secret sharing a protocol where each player receives one equal share of the encoded secret and a threshold of any k players can access the secret. This scheme is a primitive protocol by which any other secret sharing is achieved. In this work, we treated the case (k,n)=(3,5) where the dealer sends through five depolarizing channels, a quantum secret encoded in a five qubits graph state [1].

## 3.1 Introduction to Graph state

Graph states are a an efficient tool for multipartite quantum information processing task like secret sharing. Also, they have a graphical representation witch offers an intuitive picture of information flow. The secret to be shared is encoded onto classical labels placed on vertices of the graph representing local operations. The entanglement of the graph state allows these labels to be shifted around, giving us the opportunity to see graphically which set of players can access the secret [1].

## 3.2 Five qubits graph state

The five qubits graph state $|G\rangle$ given by equation (1) is schematized in figure 1 where the vertices represent the qubits and the edge the controlled-z gate. The graph state $|\Psi_G\rangle$ given by equation (2) and containing the quantum secret $|\Psi_s\rangle = \alpha|0\rangle + \beta|1\rangle = cos(\theta/2)|0\rangle + e^{i\phi}sin(\theta/2)|1\rangle$, should be transmitted by a dealer to five

players through five different channels. First, he constructs the state $|G\rangle$ from an initial five qubits state $|\Psi_0\rangle = |00000\rangle$, then applies the Hadamard gate H on each qubit and the controlled-Z gate CZ on qubits $[1,2],[2,3],[3,4],[4,5],[5,1]$ to obtain :

$$|G\rangle = \prod_{1 \leq i \leq 4} CZ_{[i,i+1]} |+\rangle^{\otimes 5} \qquad (1)$$

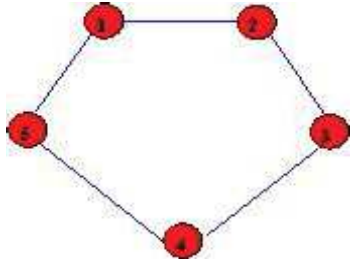

Figure 1: Five-qubits graph state $|G\rangle$.

The dealer makes an intrication between an additional qubit called D and each of the five qubits, then add to the obtained system the secret qubit S in the state $|\Psi_s\rangle = \alpha |0\rangle + \beta |1\rangle$. Then, he performs a Bell measurement on qubits D and S and obtains finally [1] :

$$|\Psi_G\rangle = \alpha |G\rangle + \beta[\prod_{1 \leq i \leq 5} Z_i] |G\rangle \qquad (2)$$

We obtain in the Dirac notation :

$|\Psi_G\rangle = (\sqrt{2}/8)\{(\alpha + \beta)[|10100\rangle + |10010\rangle + |10111\rangle + |01111\rangle + |11101\rangle + |00000\rangle + |00011\rangle + |11000\rangle + |10001\rangle + |01100\rangle + |00101\rangle + |01010\rangle] + (\alpha - \beta)[|11010\rangle + |10110\rangle + |10101 + |10000\rangle] + (-\alpha + \beta)[|01011\rangle + |01101\rangle + |00001\rangle + |10011\rangle + |01000\rangle + |11111\rangle + |00100\rangle + |01110\rangle + |00111\rangle + |11100\rangle + |11001\rangle] - (\alpha + \beta)[|11011\rangle + |00110\rangle + |01001 + |11110\rangle + |00010\rangle]\}$ \hfill (3)

## 3.3 Perfect channel

The graph state $|\Psi_G\rangle$ can be decomposed in terms of Bell states $|B_{ij}\rangle_{13}$ and $|B_{ij}\rangle_{45}$ [1] :

$|\Psi_G\rangle = (\frac{1}{2})\{|B_{00}\rangle_{13} [\alpha |+\rangle + \beta |-\rangle]_2 |B_{01}\rangle_{45} + |B_{01}\rangle_{13} [\alpha |+\rangle - \beta |-\rangle]_2 |B_{10}\rangle_{45} + |B_{10}\rangle_{13} [\alpha |-\rangle - \beta |+\rangle]_2 |B_{00}\rangle_{45} + |B_{11}\rangle_{13} [\alpha |-\rangle + \beta |+\rangle]_2 |B_{11}\rangle_{45})\}$ \hfill $(4a)$

Where the Bell states are :

$$|B_{00}\rangle = \frac{|00\rangle + |11\rangle}{\sqrt{2}}, \qquad |B_{01}\rangle = \frac{|00\rangle - |11\rangle}{\sqrt{2}},$$
$$|B_{10}\rangle = \frac{|01\rangle + |10\rangle}{\sqrt{2}}, \qquad |B_{11}\rangle = \frac{|01\rangle - |10\rangle}{\sqrt{2}}. \qquad (4b)$$

The secret should be accessible only for player 1, 2 and 3, players 4 and 5 being considered as eavesdroppers. Players 1 and 3 measure their qubits in the Bell basis and transmit the result to player 2 which applies on its qubit the suitable recovering gate R$_G$ given in table 1 to access the secret state [1].

We describe below this procedure. Equati(4a) can be written :

$$|\Psi_G\rangle = |B_{00}\rangle_{13} |\Psi_a\rangle_{245} + |B_{01}\rangle_{13} |\Psi_b\rangle_{245} + |B_{10}\rangle_{13} |\Psi_c\rangle_{245} + |B_{11}\rangle_{13} |\Psi_d\rangle_{245}$$
$$(5a)$$

Where $|\Psi_a\rangle_{245} = \frac{1}{2}\left[\alpha|+\rangle + \beta|-\rangle\right]_2 |B_{01}\rangle_{45}$ , $|\Psi_b\rangle_{245} = \frac{1}{2}\left[\alpha|+\rangle - \beta|-\rangle\right]_2 |B_{10}\rangle_{45}$ ,
$(|\Psi_c\rangle_{245} = \frac{1}{2}\left[\alpha|-\rangle - \beta|+\rangle\right]_2 |B_{00}\rangle_{45}$ , $|\Psi_d\rangle_{245} = \frac{1}{2}\left[\alpha|-\rangle + \beta|+\rangle\right]_2 |B_{11}\rangle_{45}$ ,
$$(5b)$$

Measurement in the Bell base $\{|B_{ij}\rangle_{13}\}$ gives only one term of the superposition in (5), then the global density matrix :

$$\rho_{1..5} = \frac{(|B_{ij}\rangle\langle B_{ij}|)_{13}(|\Psi\rangle_x\langle\Psi|_x)_{245}}{4} = \frac{(\rho_{ij})_{13}(\rho_x)_{245}}{4} \text{ With } x = a, b, c \text{ or } d \qquad (6)$$

The partial trace over qubits (4,5) gives the density matrix of qubit 2 :

$$\rho'_2 = \left|\Psi'_2\right\rangle \left\langle\Psi'_2\right| = P_{tr}[(\rho_x)_{245}]_{4,5} \qquad (7)$$

Then the secret state :

$$\rho_2 = R_g^* \rho'_2 R_g \quad \text{or} \quad |\Psi_2\rangle = R_g \left|\Psi'_2\right\rangle \qquad (8)$$

| $|B_{ij}\rangle_{13}$ | B$_{00}$ | B$_{01}$ | B$_{10}$ | B$_{11}$ |
|---|---|---|---|---|
| $R_g$ | H | ZH | ZXH | XH |

**Table** 1 : Secret recovering gate $R_g$ used by player 2 versus the Bell state $|B_{ij}\rangle_{13}$ measured by players 1 and 3.

## 3.4 Bit and phase flip

The sent qubits can be affected by error X, Z or Y represented respectively by the Pauli matrix $X = \begin{pmatrix} 0 & 1 \\ 1 & 0 \end{pmatrix}$ , $Z = \begin{pmatrix} 1 & 0 \\ 0 & -1 \end{pmatrix}$ and $Y = -iXZ = i \begin{pmatrix} 0 & -1 \\ 1 & 0 \end{pmatrix}$ corresponding to rotation $\pi$ around $ox$ or $oz$ or both in the block sphere. We applied the procedure described in 3.3 and obtain the table 2 giving the affected secret state $\left| \Psi^E \right\rangle_2$.

| $Error$ | $\left\| \Psi^E \right\rangle_2$ |
|---|---|
| $X_1, X_3, Y_2$ | $\alpha \left\| 1 \right\rangle - \beta \left\| 0 \right\rangle$ |
| $Z_1, X_2, Z_3$ | $\alpha \left\| 0 \right\rangle - \beta \left\| 1 \right\rangle$ |
| $Z_2, Y_1, Y_3$ | $\alpha \left\| 1 \right\rangle + \beta \left\| 0 \right\rangle$ |

**Table** 2 : Qubit 2 state $\left| \Psi^E \right\rangle_2$ versus error on qubits 1, 2 and 3.

## 3.5 Fidelity

The fidelity is one of the mathematical quantities which permits to know how close are two quantum states represented by the density matrix $\sigma$ and $\rho$ by measuring a distance between them [7] :

$$F(\sigma, \rho) = \left| Tr(\sqrt{\sqrt{\sigma}\rho\sqrt{\sigma}}) \right|^2 \qquad (9)$$

In the case of a pure state $\sigma = \left| \Psi \right\rangle \left\langle \Psi \right|$ and an arbitrary state $\rho$, the fidelity is the overlap between the two states [7] :

$$F(\left| \Psi \right\rangle, \rho) = \left\langle \Psi \right| \rho \left| \Psi \right\rangle \qquad (10)$$

In this work we measure the overlap between the correct secret state $\sigma_s = \left| \Psi_s \right\rangle \left\langle \Psi_s \right|$ and the qubit state $\rho_2 = \left| \Psi_2 \right\rangle \left\langle \Psi_2 \right|$ measured by player 2 to access the secret. Then, the fidelity is function of the angles $(\theta, \phi)$ in the Block sphere and the average fidelity is :

$$F_a = (1/4\pi) \iint F(\theta, \phi) sin(\theta) d\theta d\phi, \text{ with } 0 \leq \theta \leq \pi \text{ and } 0 \leq \phi \leq 2\pi \qquad (11)$$

We will describe below the procedure giving the fidelity. If any Pauli errors $E = \sigma_0$, $\sigma_x, \sigma_y$ or $\sigma_z$ affects the state $\left| \Psi_G \right\rangle$ in the transmission channel, we can see from equation $(4a)$ that equation $(5a)$ will keep the same form and becomes :

$$\left|\Psi_G^E\right\rangle = |B_{00}\rangle_{13}\left|\Psi_a^E\right\rangle_{245} + |B_{01}\rangle_{13}\left|\Psi_b^E\right\rangle_{245} + |B_{10}\rangle_{13}\left|\Psi_c^E\right\rangle_{245} + |B_{11}\rangle_{13}\left|\Psi_d^E\right\rangle_{245} \tag{12}$$

$\left|\Psi_{a,b,c,d}^E\right\rangle_{245}$ are the new global states of qubits system $(2,4,5)$ modified by the channel errors.

We note that $\left|\Psi_{a,b,c,d}^E\right\rangle_{245} \neq E\left|\Psi_{a,b,c,d}\right\rangle_{245}$ as channel error can affect qubits $(1,3)$ as well as qubits $(2,4,5)$. In fact, if an error affect qubits $(1,3)$ or $(4,5)$, then their global state will simply change to another Bell state. Similarly, if qubit 2 is affected, then its state will only switch to one of the forms appearing in equation $(4a)$.

After measuring on the Bell states of qubits $(1,3)$ only one term will remain in $(12)$ :

$$\left|\Psi_G^E\right\rangle' = (\tfrac{1}{2})|B_{ij}\rangle_{13}\left|\Psi_x^E\right\rangle_{245} \tag{13}$$

The corresponding affected density matrix is :

$$\rho_{1..5}^E = (\tfrac{1}{4})(|B_{ij}\rangle\langle B_{ij}|)_{13}(\left|\Psi_x^E\right\rangle\langle\Psi_x^E|)_{245} = \rho_{1..5}^E = (\tfrac{1}{4})(\rho_{ij})_{13}(\rho_x^E)_{245} \tag{14}$$

The partial trace over qubits $(4,5)$ gives the measured density matrix of qubit 2 :

$$\rho_2'^E = \left|\Psi_2'^E\right\rangle\langle\Psi_2'^E| = P_{tr}[(\rho_x^E)_{245}]_{4,5} \tag{15}$$

Then the affected secret state measured by player two :

$$\rho_2^E = R_G^*\rho_2'^E R_G = \left|\Psi_2^E\right\rangle\langle\Psi_2^E| \tag{16}$$

We multiply by the secret state $|\Psi_s\rangle = \alpha|0\rangle + \beta|1\rangle$ to obtain the fidelity :

$$F(\theta,\phi) = \langle\Psi_s|\rho_2^E|\Psi_s\rangle \tag{17}$$

Table $3a$ gives the fidelity $F(\theta,\phi)$ calculated by Feynmann Program for all the errors on qubits $i = 1, 2$ or $3$. The figure 2 shows the fidelity $F(\theta,\phi_0 = 0 \ or \ \frac{\pi}{2})$ function of the angle $\theta$ for error occurring with probability $P = 1$ on one, two and three noisy channels. We note for example that if $|\Psi_s\rangle = \frac{\sqrt{2}}{2}(|0\rangle + |1\rangle)$ the fidelity is the best $(F_a = 1)$ for error $X_1$ and the worst $(F_a = 0)$ for error $Z_1$. Also, we deduce from equation 4 that any errors on qubits 4 and 5 do not affect the secret state giving then fidelity equal to one.

| $Error$ | $|\Psi\rangle_2$ | $F(\theta,\phi)$ | $F_a$ |
|---|---|---|---|
| $\varepsilon_a$ | $\alpha|1\rangle - \beta|0\rangle$ | $|sin(\theta)sin\phi|^2$ | $1/3$ |
| $\varepsilon_b$ | $\alpha|0\rangle - \beta|1\rangle$ | $cos^2(\theta)$ | $1/3$ |
| $\varepsilon_c$ | $\alpha|1\rangle + \beta|0\rangle$ | $|sin(\theta)cos\phi|^2$ | $1/3$ |
| $\varepsilon_d$ | $\alpha|0\rangle + \beta|1\rangle$ | $1$ | $1$ |

**Table** $3a$ : Fidelity and measured state $|\Psi_2\rangle$ versus errors on qubits $i = 1, 2, 3$. The errors groups are depicted on table 3b.

| | |
|---|---|
| $\varepsilon_a$ | $X_1, X_3, Y_2, X_1X_2Z_3, Z_1X_2X_3, Y_1X_2, Y_1Z_3, Z_1Y_3, X_2Y_3,$ |
| | $Y_1Y_2Y_3, Y_1Z_2X_3, X_1Y_2X_3, X_1Z_2Y_3, Z_1Y_2Z_3$ |
| $\varepsilon_b$ | $Z_1, X_2, Z_3, X_1Z_2, Z_2X_3, X_1X_2X_3, Z_1X_2Z_3, Y_1Y_2, Y_1X_3, Y_2Y_3,$ |
| | $X_1Y_3, Y_1X_2Y_3, Y_1Z_2Z_3, X_1Y_2Z_3, Z_1Y_2X_3, Z_1Z_2Y_3$ |
| $\varepsilon_c$ | $Z_2, Y_1, Y_3, X_1X_2, X_2X_3, Z_2Z_3, X_1Z_3, Z_1Z_2, Z_1X_3, Z_1Z_2Z_3, X_1Z_2X_3,$ |
| | $Y_2Z_3, Z_1Y_2, Y_1Y_2X_3, Y_1X_2Z_3, Y_1Z_2Y_3, X_1Y_2Y_3, Z_1X_2Y_3$ |
| $\varepsilon_d$ | $X_1X_3, Z_1Z_3, Z_1X_2, X_2Z_3, X_1Z_2Z_3, Z_1Z_2X_3, Y_1Y_3, Y_1Z_2, Y_2X_3,$ |
| | $X_1Y_2, Z_2Y_3, Y_1Y_2Z_3, Y_1X_2X_3, Z_1Y_2Y_3, X_1X_2Y_3,$ |

**Table** $3b$ : Error groups with same average fidelity.

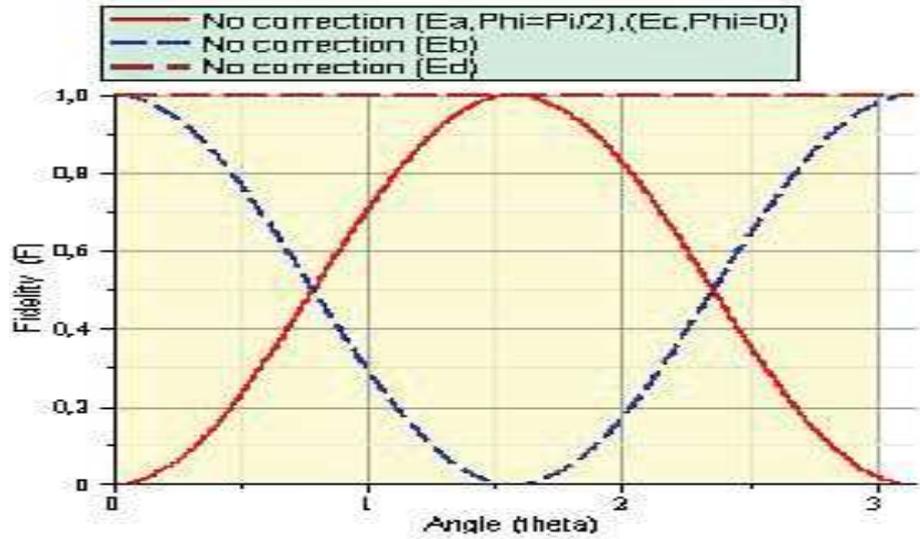

**Figure 2** : Fidelity $F(\theta,\phi)$ with $0 < \theta < \pi$, $\phi = \frac{\pi}{2}$ for errors $\varepsilon_a$ and $\phi = 0$ for errors $\varepsilon_c$. We note that for errors $\varepsilon_d$ we have $F(\theta,\phi) = 1$ for $0 < (\theta,\phi) < \pi$.

### 3.6 Depolarizing channel

The depolarizing channel is a particular model for the noise on quantum systems. In this process, the global density matrix $\rho$ is replaced by a mixed one $\rho(P)$ function of the probability $P$ that a Pauli error $E_{ij} = (\sigma_{1j} = \sigma_{xj}, \sigma_{2j} = \sigma_{yj}$ or $\sigma_{3j} = \sigma_{zj})$ affects any qubit "$j$" in the n-qubits system. For one-qubit system the matrix density is given by (18) [2] and for the n-qubits system it can be generalized by (19) :

$$\rho_1(P) = (1-P)\rho + \frac{P}{3}[X\rho X + Y\rho Y + Z\rho Z] \tag{18}$$

$$\rho_n(P) = (1-P)^n\rho + .. + \frac{P^k}{3^k}(1-P)^{n-k}[\sum_{1\le i\le 3}^{1\le j_i\le n}(\Pi_{1\le l\le k}\,\sigma^*_{ij_l})]\rho(\Pi_{1\le l\le k}\,\sigma_{ij_l}) + ...$$
$$..+\frac{P^n}{3^n}[\sum_{1\le i\le 3}^{1\le j_i\le n}(\Pi_{1\le l\le n}\,\sigma^*_{ij_l})]\rho(\Pi_{1\le l\le n}\,\sigma_{ij_l}) \tag{19}$$

Consider now the case where the five qubits are sent by the dealer through five depolarizing channels. Suppose the probability $P$ that any single error occurs on any qubit is the same in the five channels. Then we can use equation (19) as if the dealer send the five qubits through only one depolarized channel. We describe below the procedure to obtain the average fidelity $F_a(P)$, considering all the possible errors in the five transmitting noisy channels. We begin by writing the affected density matrix $\rho^E_{1..5}(P)$ received by the five players :

$$\rho^E_{1..5}(P) = (1-P)^5\rho_{1..5} + \frac{P}{3}(1-P)^4[\ \rho^{E_1}_{1..5} + \rho^{E_2}_{1..5} + \rho^{E_3}_{1..5} + \rho^{E_4}_{1..5} + \rho^{E_5}_{1..5}\ ] + \frac{P^2}{9}(1-P)^3[\rho^{E_1E_2}_{1..5} + \rho^{E_1E_3}_{1..5} + \rho^{E_1E_4}_{1..5} + \rho^{E_1E_5}_{1..5} + \rho^{E_2E_3}_{1..5} + \rho^{E_2E_4}_{1..5} + \rho^{E_2E_5}_{1..5} + \rho^{E_3E_4}_{1..5} + \rho^{E_3E_5}_{1..5} + \rho^{E_4E_5}_{1..5}] + \frac{P^3}{27}(1-P)^2[\rho^{E_1E_2E_3}_{1..5} + \rho^{E_1E_2E_4}_{1..5} + \rho^{E_1E_2E_5}_{1..5} + \rho^{E_1E_3E_4}_{1..5} + \rho^{E_1E_3E_5}_{1..5} + \rho^{E_1E_4E_5}_{1..5} + \rho^{E_2E_3E_4}_{1..5} + \rho^{E_2E_3E_5}_{1..5} + \rho^{E_2E_4E_5}_{1..5} + \rho^{E_3E_4E_5}_{1..5}] + \frac{P^4}{81}(1-P)[\rho^{E_1E_2E_3E_4}_{1..5} + \rho^{E_1E_2E_3E_5}_{1..5} + \rho^{E_1E_2E_4E_5}_2 + \rho^{E_1E_3E_4E_5}_{1..5} + \rho^{E_2E_3E_4E_5}_{1..5}] + \frac{P^5}{243}\rho^{E_1E_2E_3E_4E_5}_{1..5} \tag{20}$$

With $\rho^{E_i}_{1..5}$ the density matrix affected by errors on qubit "i" :

$$\rho^{E_i}_{1..5} = X_i\rho_{1..5}X_i + Y_i\rho_{1..5}Y_i + Z_i\rho_{1..5}Z_i \tag{21}$$

The density matrix $\rho^{E_iE_j}_{1..5}$, $\rho^{E_iE_jE_k}_{1..5}$, $\rho^{E_iE_jE_kE_l}_{1..5}$ and $\rho^{E_iE_jE_kE_lE_m}_{1..5}$ are summation of respectively 9, 27, 81 and 243 terms and represent the density matrix affected by error on two, three, four and five qubits.

After measuring on the Bell states of qubits (1,3) we obtain :

$$\rho^E_{245}(P) = (1-P)^5\rho_{245} + \frac{P}{3}(1-P)^4[\ \rho^{E_1}_{245} + \rho^{E_2}_{245} + \rho^{E_3}_{245} + \rho^{E_4}_{245} + \rho^{E_5}_{245}\ ] + \frac{P^2}{9}(1-P)^3[\rho^{E_1E_2}_{245} + \rho^{E_1E_3}_{245} + \rho^{E_1E_4}_{245} + \rho^{E_1E_5}_{245} + \rho^{E_2E_3}_{245} + \rho^{E_2E_4}_{245} + \rho^{E_2E_5}_{245} + \rho^{E_3E_4}_{245} +$$

$\rho_{245}^{E_3 E_5} + \rho_{245}^{E_4 E_5}] + \frac{P^3}{27}(1-P)^2[\rho_{245}^{E_1 E_2 E_3} + \rho_{245}^{E_1 E_2 E_4} + \rho_{245}^{E_1 E_2 E_5} + \rho_{245}^{E_1 E_3 E_4} + \rho_{245}^{E_1 E_3 E_5} + \rho_{245}^{E_1 E_4 E_5} + \rho_{245}^{E_2 E_3 E_4} + \rho_{245}^{E_2 E_3 E_5} + \rho_{245}^{E_2 E_4 E_5} + \rho_{245}^{E_3 E_4 E_5}] + \frac{P^4}{81}(1-P)[\rho_{245}^{E_1 E_2 E_3 E_4} + \rho_{245}^{E_1 E_2 E_3 E_5} + \rho_{245}^{E_1 E_2 E_4 E_5} + \rho_{245}^{E_1 E_3 E_4 E_5} + \rho_{245}^{E_2 E_3 E_4 E_5}] + \frac{P^5}{243}\rho_{245}^{E_1 E_2 E_3 E_4 E_5}$  (22)

With $\rho_{245} = (\rho_a)_{245}, (\rho_b)_{245}, (\rho_c)_{245}$ or $(\rho_d)_{245}$ and $\rho_{245}^E = (\rho_a^E)_{245}, (\rho_b^E)_{245}, (\rho_c^E)_{245}$ or $(\rho_d^E)_{245}$

After tracing over qubits $(4,5)$ and multiplying by the recovering gate $R_g$ we obtain :

$\rho_2^E(P) = (1-P)^5 \rho_2 + \frac{P}{3}(1-P)^4[\rho_2^{E_1} + \rho_2^{E_2} + \rho_2^{E_3} + \rho_2^{E_4} + \rho_2^{E_5}] + \frac{P^2}{9}(1-P)^3[\rho_2^{E_1 E_2} + \rho_2^{E_1 E_3} + \rho_2^{E_1 E_4} + \rho_2^{E_1 E_5} + \rho_2^{E_2 E_3} + \rho_2^{E_2 E_4} + \rho_2^{E_2 E_5} + \rho_2^{E_3 E_4} + \rho_2^{E_3 E_5} + \rho_2^{E_4 E_5}] + \frac{P^3}{27}(1-P)^2[\rho_2^{E_1 E_2 E_3} + \rho_2^{E_1 E_2 E_4} + \rho_2^{E_1 E_2 E_5} + \rho_2^{E_1 E_3 E_4} + \rho_2^{E_1 E_3 E_5} + \rho_2^{E_1 E_4 E_5} + \rho_2^{E_2 E_3 E_4} + \rho_2^{E_2 E_3 E_5} + \rho_2^{E_2 E_4 E_5} + \rho_2^{E_3 E_4 E_5}] + \frac{P^4}{81}(1-P)[\rho_2^{E_1 E_2 E_3 E_4} + \rho_2^{E_1 E_2 E_3 E_5} + \rho_2^{E_1 E_2 E_4 E_5} + \rho_2^{E_1 E_3 E_4 E_5} + \rho_2^{E_2 E_3 E_4 E_5}] + \frac{P^5}{243}\rho_2^{E_1 E_2 E_3 E_4 E_5}$  (23)

With $\rho_2 = |\Psi_s\rangle \langle \Psi_s|$ the correct secret and $\rho_2^E = (\rho_a^E)_2, (\rho_b^E)_2, (\rho_c^E)_2$ or $(\rho_d^E)_2$ the secret state disturbed by error $E = E_i, E_i E_j, E_i E_j E_k, E_i E_j E_k E_l,$ or $E_i E_j E_k E_l E_m$.

We multiply by the secret state and integrate over $(\theta, \phi)$ to obtain the average fidelity :

$\langle \Psi_s| \rho_2^E |\Psi_s\rangle = (1-P)^5 + \frac{P}{3}(1-P)^4[F_a^{E_1} + F_a^{E_2} + F_a^{E_3} + F_a^{E_4} + F_a^{E_5}] + \frac{P^2}{9}(1-P)^3[F_a^{E_1 E_2} + F_a^{E_1 E_3} + F_a^{E_1 E_4} + F_a^{E_1 E_5} + F_a^{E_2 E_3} + F_a^{E_2 E_4} + F_a^{E_2 E_5} + F_a^{E_3 E_4} + F_a^{E_3 E_5} + F_a^{E_4 E_5}] + \frac{P^3}{27}(1-P)^2[F_a^{E_1 E_2 E_3} + F_a^{E_1 E_2 E_4} + F_a^{E_1 E_2 E_5} + F_a^{E_1 E_3 E_4} + F_a^{E_1 E_3 E_5} + F_a^{E_1 E_4 E_5} + F_a^{E_2 E_3 E_4} + F_a^{E_2 E_3 E_5} + F_a^{E_2 E_4 E_5} + F_a^{E_3 E_4 E_5}] + \frac{P^4}{81}(1-P)[F_a^{E_1 E_2 E_3 E_4} + F_a^{E_1 E_2 E_3 E_5} + F_a^{E_1 E_2 E_4 E_5} + F_a^{E_1 E_3 E_4 E_5} + F_a^{E_2 E_3 E_4 E_5}] + \frac{P^5}{243}F_a^{E_1 E_2 E_3 E_4 E_5}$  (24)

With $\langle \Psi_s| \rho_2 |\Psi_s\rangle = 1$ and $F_a^E = \langle \Psi_s| \rho_2^E |\Psi_s\rangle$  (25)

We can write (24) as :

$\langle \Psi_s| \rho_s^E |\Psi_s\rangle = (1-P)^5 + \frac{P}{3}(1-P)^4[A] + \frac{P^2}{9}(1-P)^3[B] + \frac{P^3}{27}(1-P)^2[C] + \frac{P^4}{81}(1-P)$  (26)

We deduce from tables $3a$ and $3b$, the values of $F_a^E$ contained in table 4 :

| $F_a^E$ | Values |
|---|---|
| $F_a^{E_1}, F_a^{E_2}, F_a^{E_3}$ | $(3\times\frac{1}{3})= 1$ |
| $F_a^{E_4}, F_a^{E_5}$ | $(3\times 1) = 3$ |
| $F_a^{E_1E_2}, F_a^{E_1E_3}, F_a^{E_2E_3}$ | $(6\times\frac{1}{3})+(3\times 1) = 5$ |
| $F_a^{E_1E_4}, F_a^{E_1E_5}, F_a^{E_2E_4},$ $F_a^{E_2E_5}, F_a^{E_3E_4}, F_a^{E_3E_5}$ | $(9\times\frac{1}{3})= 3$ |
| $F_a^{E_4E_5}$ | $(9\times 1) = 9$ |
| $F_a^{E_1E_2E_3}$ | $(21\times\frac{1}{3})+ (6\times 1) = 13$ |
| $F_a^{E_1E_2E_4}, F_a^{E_1E_3E_4}, F_a^{E_2E_3E_4},$ $F_a^{E_1E_2E_5}, F_a^{E_1E_3E_5}, F_a^{E_2E_3E_5}$ | $[(6\times 3)\times\frac{1}{3}]+[(3\times 3)\times 1] = 15$ |
| $F_a^{E_1E_4E_5}, F_a^{E_2E_4E_5}, F_a^{E_3E_4E_5}$ | $(3\times 9)\times\frac{1}{3}= 9$ |
| $F_a^{E_1E_2E_3E_4}, F_a^{E_1E_2E_3E_5}$ | $[(21\times 3)\times\frac{1}{3}]+[(6\times 3)\times 1] = 39$ |
| $F_a^{E_1E_2E_4E_5}, F_a^{E_1E_3E_4E_5}, F_a^{E_2E_3E_4E_5}$ | $(6\times 9)\times\frac{1}{3}+(3\times 9)\times 1 = 45$ |
| $F_a^{E_1E_2E_3E_4E_5}$ | $(21\times 9)\times\frac{1}{3}+(6\times 9)\times 1 = 117$ |

**Table** 4 : Values of $F_a^E$ for all the possible errors.

The calculus gives :

$$A = 9, \quad B = 42, \quad C = 130, \quad D = 213, \quad E = 117 \tag{27}$$

The average fidelity is then :

$$F_a(P) = (1 - P)^5 + 3P(1 - P)^4 + \tfrac{14}{3}P^2(1 - P)^3 + \tfrac{130}{27}P^3(1 - P)^2 + \tfrac{71}{27}P^4(1 - P) + \tfrac{13}{27}P^5 \tag{28}$$

Finally :

$$F_a(P) = 1 - 2P + \tfrac{8}{3}P^2 - \tfrac{32}{27}P^3 \tag{29}$$

# 4. The five-qubits code

We describe below the channel error correction by the five qubits code. This code is described in [7][8] and uses five qubits to protect one of them in a superposed state from any error X, Y or Z. We note that when using any of the three codes, the fidelity is always equal to one if only one among the five sents qubits is disturbed in the transmiting channel. However, if two sent qubits and more are disturbed during the transmission, then the fidelity will vary upon the used code.

## 4.1 Bit and phase flip

The dealer protects each of the three qubits $(1, 2, 3)$ with four ancillas as showed in figure 3 and sends them through three noisy channel which introduces a bit or phase flip or both with probability $P = 1$.

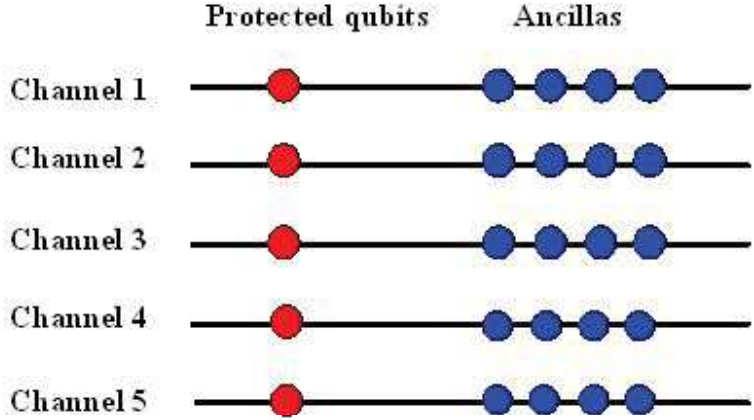

**Figure** $3$ : Transmission of five protected qubits through five channels. We note that in general the dealer will protect the five qubits as he don't know which players will access the secret.

The graph state (3) can be written as follow :

$$|\Psi_G\rangle = (\sqrt{2}/8)\{[\alpha\,|0\rangle + \beta\,|1\rangle]_i\,|\Psi_a\rangle_{jklm} + [\alpha\,|0\rangle - \beta\,|1\rangle]_i\,|\Psi_b\rangle_{jklm} + [\beta\,|0\rangle + \alpha\,|1\rangle]_i\,|\Psi_c\rangle_{jklm} + [\beta\,|0\rangle - \alpha\,|1\rangle]_i\,|\Psi_d\rangle_{jklm}] \quad (30)$$

With $[i = 1, (\ j, k, l, m) = 2, 3, 4, 5], [i = 2, (j, k, l, m) = 1, 3, 4, 5]$ or $[i = 3, (j, k, l, m) = 1, 2, 4, 5]$ $\quad (31)$

The states $|\Psi_a\rangle_{jklm}, |\Psi_b\rangle_{jklm}, |\Psi_c\rangle_{jklm}, |\Psi_d\rangle_{jklm}$ have different expressions depending on the value of "$i$". We suppose that the dealer knows that the secret will be accessed by players (1,2,3). Then he will not protect the qubits sent to players $(4, 5)$ as the channel errors on them do not affect the accessed secret. To protect qubit "i" the dealer adds four ancillas each one in the state $|0\rangle$. After applying the coding circuit we obtain the coded graph sate $|\Psi_{G_1}\rangle$ sent by the dealer, where $|0_l\rangle$ and $|1_l\rangle$ are the logical qubits [7] :

$|\Psi_{G_1}\rangle = (\sqrt{2}/8)\{[\alpha\,|0_l\rangle + \beta\,|1_l\rangle]_i\,|\Psi_a\rangle_{jklm} + [\alpha\,|0_l\rangle - \beta\,|1_l\rangle]_i\,|\Psi_b\rangle_{jklm} + [\beta\,|0_l\rangle +$
$\alpha\,|1_l\rangle]_i\,|\Psi_c\rangle_{jklm} + [\beta\,|0_l\rangle - \alpha\,|1_l\rangle]_i\,|\Psi_d\rangle_{jklm}]$ \hfill (32)

After syndrome measurement, correction and decoding, the players suppress the ancillas. If the qubit "$i$" is affected by error $E_i = X_i$ , $Y_i$ or $Z_i$ then the graph state becomes :

$\left|\Psi_G^{E_i}\right\rangle = E_i\,|\Psi_G\rangle = (\sqrt{2}/8)\{E_i[\alpha\,|0\rangle + \beta\,|1\rangle]_i\,|\Psi_a\rangle_{jklm} + E_i[\alpha\,|0\rangle - \beta\,|1\rangle]_i\,|\Psi_b\rangle_{jklm} +$
$E_i[\beta\,|0\rangle + \alpha\,|1\rangle]_i\,|\Psi_c\rangle_{jklm} + E_i[\beta\,|0\rangle - \alpha\,|1\rangle]_i\,|\Psi_d\rangle_{jklm}]$ \hfill (33)

Table 5 gives the syndromes $S$ for the five qubits code of single and double errors occurring in the logical qubit "$i$". The double errors are corrected as the single error having same syndrome. The third column gives the error $E_i$ affecting the to be protected physical qubit "$i$" after decoding obtained by Feynman Program.

| $Error$ | $S$ | $E_i$ |
|---|---|---|
| $X_i, (Z_{a_2}Z_{a_3}), (X_{a_3}Z_{a_4}, Z_{a_1}X_{a_2})$ | 0101 | $I_i, (X_i), (Z_i)$ |
| $X_{a_1}, (Z_{a_3}Z_{a_4}), (Z_iX_{a_4}, Z_{a_2}X_{a_3})$ | 0010 | $I_i, (X_i), (Z_i)$ |
| $X_{a_2}, (Z_iZ_{a_4}), (X_iZ_{a_1}, Z_{a_3}X_{a_4})$ | 1001 | $I_i, (X_i), (Z_i)$ |
| $X_{a_3}, (Z_iZ_{a_1}), (X_iZ_{a_4}, X_{a_1}Z_{a_2})$ | 0100 | $I_i, (X_i), (Z_i)$ |
| $X_{a_4}, (Z_{a_1}Z_{a_2}), (X_{a_2}Z_{a_3}, Z_iX_{a_1})$ | 1010 | $I_i, (X_i), (Z_i)$ |
| $Z_i, (X_{a_1}X_{a_4}), (X_{a_2}Z_{a_4}, Z_{a_1}X_{a_3})$ | 1000 | $I_i, (X_i), (Z_i)$ |
| $Z_{a_1}, (X_iX_{a_2}), (Z_iX_{a_3}, Z_{a_2}X_{a_4})$ | 1100 | $I_i, (X_i), (Z_i)$ |
| $Z_{a_2}, (X_{a_1}X_{a_3}), (X_iZ_{a_3}, Z_{a_1}X_{a_4})$ | 0110 | $I_i, (X_i), (Z_i)$ |
| $Z_{a_3}, (X_{a_2}X_{a_4}), (X_iZ_{a_2}, X_{a_1}Z_{a_4})$ | 0011 | $I_i, (X_i), (Z_i)$ |
| $Z_{a_4}, (X_iX_{a_3}), (X_{a_1}Z_{a_3}, Z_iX_{a_2})$ | 0001 | $I_i, (X_i), (Z_i)$ |
| $Y_i, (X_{a_2}X_{a_3}, Z_{a_1}Z_{a_4})$ | 1101 | $I_i, (Y_i)$ |
| $Y_{a_1}, (X_{a_3}X_{a_4}, Z_iZ_{a_2})$ | 1110 | $I_i, (Y_i)$ |
| $Y_{a_2}, (X_iX_{a_4}, Z_{a_1}Z_{a_3})$ | 1111 | $I_i, (Y_i)$ |
| $Y_{a_3}, (X_iX_{a_1}, Z_{a_2}Z_{a_4})$ | 0111 | $I_i, (Y_i)$ |
| $Y_{a_4}, (X_{a_1}X_{a_2}, Z_iZ_{a_3})$ | 1011 | $I_i, (Y_i)$ |

**Table** 5 : The error $E_i$ affecting the physical qubit "i" versus errors on the logical qubit "i". The ancillas are designed by "$a_j$" and the to be protected qubit by "$i$" with $i = 1, 2, 3$ and $j = 1, 2, 3, 4$.

## 4.2 Depolarizing channel

Consider three double errors $E_kE_l$, $E_mE_n$ and $E_oE_p$ occurring respectively in channels "1", "2" and "3" on any qubits (k,l,m,n,o,p) and corrected as the three single error with similar syndrome. The qubits $(4,5)$ are not protected and can be affected by any error $X$, $Y$ or $Z$. If the probability that channel error occurs on one qubit is equal to $P$, then the density matrix received by the five players is :

$$\rho^E_{(123a)(45)} = (1-P)^8 \rho_{(123a)(45)} + \frac{P}{3}(1-P)^7 [\sum \rho^{E_x}_{(123a)(45)}] + \frac{P^2}{3^2}(1-P)^6 [\sum \rho^{E_x E_y}_{(123a)(45)}] +$$
$$\frac{P^3}{3^3}(1-P)^5 [\sum \rho^{E_x E_y E_z}_{(123a)(45)}] + \frac{P^4}{3^4}(1-P)^4 [\sum \rho^{E_x E_y E_z E_u}_{(123a)(45)}] + \frac{P^5}{3^5}(1-P)^3 [\sum \rho^{E_x E_y E_z E_u E_v}_{(123a)(45)}] +$$
$$\frac{P^6}{3^6}(1-P)^2 [\sum \rho^{E_x E_y E_z E_u E_v E_w}_{(123a)(45)}] + \frac{P^7}{3^7}(1-P) [\sum \rho^{E_x E_y E_z E_u E_v E_w E_s}_{(123a)(45)}] + \frac{P^8}{3^8} [\sum \rho^{E_k E_l E_m E_n E_o E_p E_4 E_5}_{(123a)(45)}]$$

$$(34)$$

The notation $(123a)$ represents the logical qubits 1,2 and 3, each one protected by four ancillas and :

$$\sum \rho^{E_x}_{(123a)(45)} = \rho^{E_k}_{(123a)(45)} + \rho^{E_l}_{(123a)(45)} + \rho^{E_m}_{(123a)(45)} + \rho^{E_n}_{(123a)(45)} + \rho^{E_o}_{(123a)(45)} +$$
$$\rho^{E_p}_{(123a)(45)} + \rho^{E_4}_{(123a)(45)} + \rho^{E_5}_{(123a)(45)} \qquad (35a)$$

$$\rho^{E_k}_{(123a)(45)} = \rho^{X_k}_{(123a)(45)} + \rho^{Y_k}_{(123a)(45)} + \rho^{Z_k}_{(123a)(45)} \;;\; \rho^{X_k}_{(123a)(45)} = X_k^* \rho_{(123a)(45)} X_k \quad (35b)$$

The summations $\sum \rho^{E_x}_{(123a)(45)}$, $\sum \rho^{E_x E_y}_{(123a)(45)}$, $\sum \rho^{E_x E_y E_z}_{(123a)(45)}$, $\sum \rho^{E_x E_y E_z E_u}_{(123a)(45)}$, $\sum \rho^{E_x E_y E_z E_u E_v}_{(123a)(45)}$, $\sum \rho^{E_x E_y E_z E_u E_v E_w}_{(123a)(45)}$ and $\sum \rho^{E_x E_y E_z E_u E_v E_w E_s}_{(123a)(45)}$ contains respectively 8X3, 28X3$^2$, 56X3$^3$,70X3$^4$, 56X3$^5$, 28X3$^6$ and 8X3$^7$ terms. The expression $\rho^{E_k E_l E_m E_n E_o E_p E_4 E_5}_{(123a)(45)}$ is the summation of 3$^8$ terms.

After decoding and suppressing the ancillas we obtain by using tables 5 the affected graph state :

$$\rho^E_{(1..5)} = [(1-P)^8 + 18\frac{P}{3}(1-P)^7 + 108\frac{P^2}{9}(1-P)^6 + 216\frac{P^3}{27}(1-P)^5]\rho_{(1..5)} + [3\frac{P^2}{9}(1-P)^6 + 36\frac{P^3}{27}(1-P)^5 + 108\frac{P^4}{81}(1-P)^4][\rho^{E_1}_{(1..5)} + \rho^{E_2}_{(1..5)} + \rho^{E_3}_{(1..5)}] + [9\frac{P^4}{81}(1-P)^4 + 54\frac{P^5}{243}(1-P)^3][\rho^{E_1 E_2}_{(1..5)} + \rho^{E_1 E_3}_{(1..5)} + \rho^{E_2 E_3}_{(1..5)}] + [\frac{1}{27}P^6(1-P)^2][\rho^{E_1 E_2 E_3}_{(1..5)}] + [\frac{P}{3}(1-P)^7 + 18\frac{P^2}{9}(1-P)^6 + 108\frac{P^3}{27}(1-P)^5 + 216\frac{P^4}{81}(1-P)^4][\rho^{E_4}_{(1..5)} + \rho^{E_5}_{(1..5)}] + [\frac{P^2}{9}(1-P)^6 + 18\frac{P^3}{27}(1-P)^5 + 108\frac{P^4}{81}(1-P)^4 + 216\frac{P^5}{243}(1-P)^3][\rho^{E_4 E_5}_{(1..5)}] + [3\frac{P^3}{27}(1-P)^5 + 36\frac{P^4}{81}(1-P)^4 + 108\frac{P^5}{243}(1-P)^3][\rho^{E_1 E_4}_{(1..5)} + \rho^{E_1 E_5}_{(1..5)} + \rho^{E_2 E_4}_{(1..5)} + \rho^{E_2 E_5}_{(1..5)} + \rho^{E_3 E_4}_{(1..5)} + \rho^{E_3 E_5}_{(1..5)}] + [3\frac{P^4}{81}(1-P)^4 + 36\frac{P^5}{243}(1-P)^3 + 108\frac{P^6}{3^6}(1-P)^2][\rho^{E_1 E_4 E_5}_{(1..5)} + \rho^{E_2 E_4 E_5}_{(1..5)} + \rho^{E_3 E_4 E_5}_{(1..5)}] + [9\frac{P^5}{243}(1-P)^3 + 54\frac{P^6}{3^6}(1-P)^2][\rho^{E_1 E_2 E_4}_{(1..5)} + \rho^{E_1 E_2 E_5}_{(1..5)} + \rho^{E_1 E_3 E_4}_{(1..5)} + \rho^{E_1 E_3 E_5}_{(1..5)} + \rho^{E_2 E_3 E_4}_{(1..5)} + \rho^{E_2 E_3 E_5}_{(1..5)}] + [9\frac{P^6}{3^6}(1-P)^2 + 54\frac{P^7}{3^7}(1-P)][\rho^{E_1 E_2 E_4 E_5}_{(1..5)} + \rho^{E_1 E_3 E_4 E_5}_{(1..5)} + \rho^{E_2 E_3 E_4 E_5}_{(1..5)}] + [27\frac{P^7}{3^7}(1-P)][\rho^{E_1 E_2 E_3 E_4}_{(1..5)} + \rho^{E_1 E_2 E_3 E_5}_{(1..5)}] + \frac{P^8}{3^8}[27\rho^{E_1 E_2 E_3 E_4 E_5}_{(1..5)}]$$

$$(36)$$

After measuring on the Bell states of qubits (1,3), tracing over qubits (4,5), multiplying by the recovering gate and the secret state and integrating we obtain the average fidelity :

$$F_a(P) = \langle \Psi_s | \rho_2^E | \Psi_s \rangle = [(1-P)^8 + 18\frac{P}{3}(1-P)^7 + 108\frac{P^2}{9}(1-P)^6 + 216\frac{P^3}{27}(1-P)^5] +$$
$$[3\frac{P^2}{9}(1-P)^6 + 36\frac{P^3}{27}(1-P)^5 + 108\frac{P^4}{81}(1-P)^4][F_a^{E_1} + F_a^{E_2} + F_a^{E_3}] + [9\frac{P^4}{81}(1-P)^4 +$$
$$54\frac{P^5}{243}(1-P)^3][F_a^{E_1 E_2} + F_a^{E_1 E_3} + F_a^{E_2 E_3}] + [\frac{1}{27}P^6(1-P)^2][F_a^{E_1 E_2 E_3}][\frac{P}{3}(1-P)^7 +$$
$$18\frac{P^2}{9}(1-P)^6 + 108\frac{P^3}{27}(1-P)^5 + 216\frac{P^4}{81}(1-P)^4][F_a^{E_4} + F_a^{E_5}] + [\frac{P^2}{9}(1-$$
$$P)^6 + 18\frac{P^3}{27}(1-P)^5 + 108\frac{P^4}{81}(1-P)^4 + 216\frac{P^5}{243}(1-P)^3][F_a^{E_4 E_5}] + [3\frac{P^3}{27}(1-$$
$$P)^5 + 36\frac{P^4}{81}(1-P)^4 + 108\frac{P^5}{243}(1-P)^3][F_a^{E_1 E_4} + F_a^{E_1 E_5} + F_a^{E_2 E_4} + F_a^{E_2 E_5} +$$
$$F_a^{E_3 E_4} + F_a^{E_3 E_5}] + [3\frac{P^4}{81}(1-P)^4 + 36\frac{P^5}{243}(1-P)^3 + 108\frac{P^6}{36}(1-P)^2][F_a^{E_1 E_4 E_5} +$$
$$F_a^{E_2 E_4 E_5} + F_a^{E_3 E_4 E_5}] + [9\frac{P^5}{243}(1-P)^3 + 54\frac{P^6}{36}(1-P)^2][F_a^{E_1 E_4 E_4} + F_a^{E_1 E_2 E_5} +$$
$$F_a^{E_1 E_3 E_4} + F_a^{E_1 E_3 E_5} + F_a^{E_2 E_3 E_4} + F_a^{E_2 E_3 E_5}] + [9\frac{P^6}{36}(1-P)^2 + 54\frac{P^7}{37}(1-P)][F_a^{E_1 E_2 E_4 E_5} +$$
$$F_a^{E_1 E_3 E_4 E_5} + F_a^{E_2 E_3 E_4 E_5}] + [27\frac{P^7}{37}(1-P)][F_a^{E_1 E_2 E_3 E_4} + F_a^{E_1 E_2 E_3 E_5}] +$$
$$27\frac{P^8}{38}[F_a^{E_1 E_2 E_3 E_4 E_5}] \tag{37}$$

We deduce from table 4 :

$$F_a(P) = [(1-P)^8 + 18\frac{P}{3}(1-P)^7 + 108\frac{P^2}{9}(1-P)^6 + 216\frac{P^3}{27}(1-P)^5] +$$
$$[3\frac{P^2}{9}(1-P)^6 + 36\frac{P^3}{27}(1-P)^5 + 108\frac{P^4}{81}(1-P)^4][3] + [9\frac{P^4}{81}(1-P)^4 +$$
$$54\frac{P^5}{243}(1-P)^3][15] + [\frac{1}{27}P^6(1-P)^2][13] + [\frac{P}{3}(1-P)^7 + 18\frac{P^2}{9}(1-P)^6 + 108\frac{P^3}{27}(1-$$
$$P)^5 + 216\frac{P^4}{81}(1-P)^4][6] + [\frac{P^2}{9}(1-P)^6 + 18\frac{P^3}{27}(1-P)^5 + 108\frac{P^4}{81}(1-P)^4 +$$
$$216\frac{P^5}{243}(1-P)^3][9] + [3\frac{P^3}{27}(1-P)^5 + 36\frac{P^4}{81}(1-P)^4 + 108\frac{P^5}{243}(1-P)^3][18] + [3\frac{P^4}{81}(1-$$
$$P)^4 + 36\frac{P^5}{243}(1-P)^3 + 108\frac{P^6}{36}(1-P)^2][27] + [9\frac{P^5}{243}(1-P)^3 + 54\frac{P^6}{36}(1-P)^2][90] +$$
$$[9\frac{P^6}{36}(1-P)^2 + 54\frac{P^7}{37}(1-P)][135] + [27\frac{P^7}{37}(1-P)][78] + 27\frac{P^8}{38}[117] \tag{38}$$

Then :

$$F_a(P) = (1-P)^8 + 8P(1-P)^7 + 26P^2(1-P)^6 + 44P^3(1-P)^5 + \frac{128}{3}P^4(1-P)^4 +$$
$$\frac{80}{3}P^5(1-P)^3 + \frac{346}{27}P^6(1-P)^2 + \frac{116}{27}P^7(1-P) + \frac{13}{27}P^8 \tag{39}$$

## 5. The seven-qubits code

This code is described in [9] and uses five qubits to protect one of them against an X, Y or Z error. Tables 6 depict the error $E_i$ on the protected qubit "i" (after correction) for different channel errors E.

| $E$ | $S$ | $E_i$ | $E$ | $S$ | $E_i$ |
|---|---|---|---|---|---|
| $X_1,(X_2X_3,X_4X_5,X_6X_7)$ | 000001 | $I_i,(X_i)$ | $Y_1$ | 001001 | $I_i$ |
| $X_2,(X_1X_3,X_4X_6,X_5X_7)$ | 000010 | $I_i,(X_i)$ | $Y_2$ | 010010 | $I_i$ |
| $X_3,(X_1X_2,X_5X_6,X_4X_7)$ | 000011 | $I_i,(X_i)$ | $Y_3$ | 011011 | $I_i$ |
| $X_4,(X_1X_5,X_2X_6,X_3X_7)$ | 000100 | $I_i,(X_i)$ | $Y_4$ | 100100 | $I_i$ |
| $X_5,(X_1X_4,X_2X_7,X_3X_6)$ | 000101 | $I_i,(X_i)$ | $Y_5$ | 101101 | $I_i$ |
| $X_6,(X_1X_7,X_2X_4,X_3X_5)$ | 000110 | $I_i,(X_i)$ | $Y_6$ | 110110 | $I_i$ |
| $X_7,(X_1X_6,X_2X_5,X_3X_4)$ | 000111 | $I_i,(X_i)$ | $Y_7$ | 111111 | $I_i$ |

**Table** 6$a$ : Syndromes and the error $E_i$ affecting the protected qubits "i" after correction by an operators $X_k$ or $Y_k$ ($k = 1..7$).

| $Errors$ | $S$ | $E_i$ |
|---|---|---|
| $Z_1,(Z_6Z_7,Z_2Z_3,Z_4Z_5)$ | 001000 | $I_i,(Z_i)$ |
| $Z_2,(Z_1Z_3,Z_4Z_6,Z_5Z_7)$ | 010000 | $I_i,(Z_i)$ |
| $Z_3,(Z_4Z_7,Z_1Z_2,Z_5Z_6)$ | 011000 | $I_i,(Z_i)$ |
| $Z_4,(Z_3Z_7,Z_1Z_5,Z_2Z_6)$ | 100000 | $I_i,(Z_i)$ |
| $Z_5,(Z_2Z_7,Z_1Z_4,Z_3Z_6)$ | 101000 | $I_i,(Z_i)$ |
| $Z_6,(Z_1Z_7,Z_2Z_4,Z_3Z_5)$ | 110000 | $I_i,(Z_i)$ |
| $Z_7,(Z_1Z_6,Z_2Z_5,Z_3Z_4)$ | 111000 | $I_i,(Z_i)$ |

**Table** 6$b$: Syndromes and the error affecting the protected qubits after correction by an operator $Z_k$ ($k = 1..7$).

| $Errors$ | $S$ | $Errors$ | $S$ | $Errors$ | $S$ |
|---|---|---|---|---|---|
| $X_1Z_4$ | 100001 | $Z_1X_4$ | 001100 | $Z_1X_3$ | 001011 |
| $X_1Z_5$ | 101001 | $Z_2X_4$ | 010100 | $X_1Z_2$ | 010001 |
| $X_1Z_6$ | 110001 | $Z_3X_4$ | 011100 | $X_1Z_3$ | 011001 |
| $X_1Z_7$ | 111001 | $Z_1X_5$ | 001101 | $X_2Z_1$ | 001010 |
| $X_2Z_7$ | 111010 | $Z_2X_5$ | 010101 | $X_2Z_3$ | 011010 |
| $X_3Z_4$ | 100011 | $Z_1X_6$ | 001110 | $X_2Z_4$ | 100010 |
| $X_3Z_6$ | 110011 | $Z_2X_6$ | 010110 | $X_2Z_5$ | 101010 |
| $X_3Z_7$ | 111011 | $Z_3X_6$ | 011110 | $X_2Z_6$ | 110010 |
| $X_4Z_7$ | 111100 | $Z_1X_7$ | 001111 | $Z_2X_3$ | 010011 |
| $X_5Z_7$ | 111101 | $Z_2X_7$ | 010111 | $X_4Z_5$ | 101100 |
| $X_6Z_5$ | 101110 | $Z_3X_7$ | 011111 | $X_4Z_6$ | 110100 |
| $X_6Z_7$ | 111110 | $Z_4X_5$ | 100101 | $Z_4X_6$ | 100110 |
| $X_5Z_6$ | 110101 | $Z_3X_5$ | 011101 | $X_3Z_5$ | 101011 |

**Table** 6$c$: Syndromes of double channels errors $Z_kX_l$ not affecting the protected qubits after correction.

## 6. The nine-qubits code

This code called the Shor code and explained in [7] uses nine qubits to protect one of them in a superposed state from any error X, Y or Z. The simulation with Feynman Program gives in tables 7 the error $E_i$ affecting the protected qubits (after correction) for different single and double channels errors $E$ on logical qubit "i".

| $E$ | $S$ | $E_i$ |
|---|---|---|
| $X_1, X_2X_3$ | 10000000 | $I_i, Z_i$ |
| $X_2, X_1X_3$ | 11000000 | $I_i, Z_i$ |
| $X_3, X_1X_2$ | 01000000 | $I_i, Z_i$ |
| $X_4, X_5X_6$ | 00100000 | $I_i, Z_i$ |
| $X_5, X_4X_6$ | 00110000 | $I_i, Z_i$ |
| $X_6, X_4X_5$ | 00010000 | $I_i, Z_i$ |
| $X_7, X_8X_9$ | 00001000 | $I_i, Z_i$ |
| $X_8, X_7X_9$ | 00001100 | $I_i, Z_i$ |
| $X_9, X_7X_8$ | 00000100 | $I_i, Z_i$ |

**Table** $7a$ : Syndromes $S$ and the $E_i$ error affecting the protected qubits after correction by an operators $X_k$ $(k = 1..9)$.

| $E$ | $S$ | $E_i$ |
|---|---|---|
| $Y_1, X_1Z_{2,3}$ | 10000010 | $I_i$ |
| $Y_2, X_2Z_{1,3}$ | 11000010 | $I_i$ |
| $Y_3, X_3Z_{1,2}$ | 01000010 | $I_i$ |
| $Y_4, X_4Z_{5,6}$ | 00100011 | $I_i$ |
| $Y_5, X_5Z_{4,6}$ | 00110011 | $I_i$ |
| $Y_6, X_6Z_{4,5}$ | 00010011 | $I_i$ |
| $Y_7, X_7Z_{8,9}$ | 00001001 | $I_i$ |
| $Y_8, X_8Z_{7,9}$ | 00001101 | $I_i$ |
| $Y_9, X_9Z_{7,8}$ | 00000101 | $I_i$ |

**Table** $7b$ : Syndromes $S$ and the $E_i$ affecting the protected qubits after correction by an operators $Y_k$ $(k = 1..9)$.

| $Errors$ | | $S$ | $E_i$ |
|---|---|---|---|
| $(Z_1, Z_2, Z_3), (Z_4Z_{7,8,9}, Z_5Z_{7,8,9}, Z_6Z_{7,8,9})$ | | 00000010 | $(I_i), (X_i)$ |
| $(Z_4, Z_5, Z_6), (Z_1Z_{7,8,9}, Z_2Z_{7,8,9}, Z_3Z_{7,8,9})$ | | 00000011 | $(I_i), (X_i)$ |
| $(Z_7, Z_8, Z_9), (Z_1Z_{4,5,6}, Z_2Z_{4,5,6}, Z_3Z_{4,5,6})$ | | 00000001 | $(I_i), (X_i)$ |
| $(Z_1Z_{2,3}, Z_2Z_3, Z_4Z_{5,6}, Z_5Z_6, Z_7Z_{8,9}, Z_8Z_9)$ | | 00000000 | $(I_i)$ |

**Table** $7c$ : Syndromes $S$ and the $E_i$ affecting the protected qubits after correction by an operator $Z_k$ $(k = 1..9)$.

| E | S | E | S | E | S |
|---|---|---|---|---|---|
| $X_1X_4$ | 10100000 | $X_2X_7$ | 11001000 | $X_4X_7$ | 00101000 |
| $X_1X_5$ | 10110000 | $X_2X_8$ | 11001100 | $X_4X_8$ | 00101100 |
| $X_1X_6$ | 10010000 | $X_2X_9$ | 11000100 | $X_4X_9$ | 00100100 |
| $X_1X_7$ | 10001000 | $X_3X_4$ | 01100000 | $X_5X_7$ | 00111000 |
| $X_1X_8$ | 10001100 | $X_3X_5$ | 01110000 | $X_5X_8$ | 00111100 |
| $X_1X_9$ | 10000100 | $X_3X_6$ | 01010000 | $X_5X_9$ | 00110100 |
| $X_2X_4$ | 11100000 | $X_3X_7$ | 01001000 | $X_6X_7$ | 00011000 |
| $X_2X_5$ | 11110000 | $X_3X_8$ | 01001100 | $X_6X_8$ | 00011100 |
| $X_2X_6$ | 11010000 | $X_3X_9$ | 01000100 | $X_6X_9$ | 00010100 |

**Table** $7d$ : Syndromes $S$ of double channels errors $X_kX_l$ not affecting the protected qubits after correction.

| E | S | E | S |
|---|---|---|---|
| $X_1Z_{4,5,6}$ | 10000011 | $X_4Z_{1,2,3}$ | 00100010 |
| $X_1Z_{7,8,9}$ | 10000001 | $X_5Z_{1,2,3}$ | 00110010 |
| $X_2Z_{7,8,9}$ | 11000001 | $X_6Z_{1,2,3}$ | 00010010 |
| $X_2Z_{4,5,6}$ | 11000011 | $X_7Z_{1,2,3}$ | 00001010 |
| $X_3Z_{4,5,6}$ | 01000011 | $X_8Z_{1,2,3}$ | 00001110 |
| $X_3Z_{7,8,9}$ | 01000001 | $X_9Z_{1,2,3}$ | 00000110 |
| $X_4Z_{7,8,9}$ | 00100001 | $X_7Z_{4,5,6}$ | 00001011 |
| $X_5Z_{7,8,9}$ | 00110001 | $X_8Z_{4,5,6}$ | 00001111 |
| $X_6Z_{7,8,9}$ | 00010001 | $X_9Z_{4,5,6}$ | 00000111 |

**Table** $7e$ : Syndromes of double channels errors $X_kZ_l$ not affecting the protected qubits after correction.

## 7. Comparison among the three codes

The procedure giving the average fidelity described in section 4 is the same for the seven and nine qubits codes. The difference comes from the number $n$ of double channel errors having an exclusive syndrome allowing their recovery, then letting the protected qubit error free. We suppose that triple channel errors and more are very unlikely. We deduce from tables 5, 6 and 7 and for each code the next fractions of recoverable double channel errors :

$$\frac{n_5}{N_5} = 0 \quad ; \quad \frac{n_7}{N_7} = \frac{39}{81} \quad ; \quad \frac{n_9}{N_9} = \frac{108}{144} \tag{40}$$

With $N_5{=}40$   $N_7{=}81$ and $N_9{=}144$ the total number of double errors for each code. We can deduce the average fidelity by changing the value $F_a = \frac{1}{3}$ in table $3a$ by an average value $f_n$  and obtain for each code :

$$f_5 = \tfrac{1}{3} \; ; \quad f_7 = \frac{(n_7 \times 1 + (N_7 - n_7) \times \frac{1}{3})}{81} = \frac{53}{81} \; ; \quad f_9 = \frac{(n_9 \times 1 + (N_9 - n_9) \times \frac{1}{3})}{144} = \frac{5}{6} \tag{41}$$

We substitute in table 4 the value $\frac{1}{3}$ by the average value $f_n$(except in the last arrow as for P=1 the errors are unrecoverable whatever is the code). Equation (39) becomes :

$F_a(P) = \langle\Psi_s|\,\rho_2^E\,|\Psi_s\rangle = [(1-P)^8 + 18\frac{P}{3}(1-P)^7 + 108\frac{P^2}{9}(1-P)^6 + 216\frac{P^3}{27}(1-P)^5] + [3\frac{P^2}{9}(1-P)^6 + 36\frac{P^3}{27}(1-P)^5 + 108\frac{P^4}{81}(1-P)^4][9f_n] + [9\frac{P^4}{81}(1-P)^4 + 54\frac{P^5}{243}(1-P)^3][3(6f_n+3)] + [-\frac{1}{27}P^6(1-P)^2][21f_n+6][\frac{P}{3}(1-P)^7 + 18\frac{P^2}{9}(1-P)^6 + 108\frac{P^3}{27}(1-P)^5 + 216\frac{P^4}{81}(1-P)^4][6] + [\frac{P^2}{9}(1-P)^6 + 18\frac{P^3}{27}(1-P)^5 + 108\frac{P^4}{81}(1-P)^4 + 216\frac{P^5}{243}(1-P)^3][\ 9] + [3\frac{P^3}{27}(1-P)^5 + 36\frac{P^4}{81}(1-P)^4 + 108\frac{P^5}{243}(1-P)^3][54f_n] + [3\frac{P^4}{81}(1-P)^4 + 36\frac{P^5}{243}(1-P)^3 + 108\frac{P^6}{3^6}(1-P)^2][81f_n] + [9\frac{P^5}{243}(1-P)^3 + 54\frac{P^6}{3^6}(1-P)^2][54(2f_n+1)] + [9\frac{P^6}{3^9}(1-P)^2 + 54\frac{P^7}{3^7}(1-P)][81(2f_n+1) + [27\frac{P^7}{3^7}(1-P)][18(7f_n+2)] + 27\frac{P^8}{3^8}[117]$

$\hfill(42)$

We obtain :

$F_a(P) = (1-P)^8 + 8P(1-P)^7 + (3f_n+25)P^2(1-P)^6 + (18f_n+38)P^3(1-P)^5 + (41f_n+29)P^4(1-P)^4 + (44f_n+12)P^5(1-P)^3 + (\frac{205f_n+47}{9})P^6(1-P)^2 + (\frac{50f_n+22}{9})P^7(1-P) + \frac{13}{27}P^8$

$\hfill(43)$

## 8. Summary

Table 8 summarizes all the results and figure 4 compares the average fidelity without and with correction by the three codes. The figure 1 shows logically that the average fidelity is decreasing with P without and with correction by any code. The values of fidelity are always better and the decrease is slower when using codes. The best average fidelity is given by the nine qubits code, followed by the seven qubits then the five qubits code. The reason is that for the five qubits code all the double errors (in logical qubit) let the protected qubits affected, while some of them could be covered when using the two other codes. We considered in this work that triple errors (in logical qubits) and more are very unlikely, so that syndrome measurement allows (in seven and nine qubits code) recovering errors. We note that if P=1, then the average fidelity $Fa(P=1)=\frac{13}{27}=0.4815$ is the same regardless the used code.

| Code | $F_a(P)$ |
|---|---|
| $C_0$ | $1 - 2P + \frac{8}{3}P^2 - \frac{32}{27}P^3$ |
| $C_n$ | $(1-P)^8 + 8P(1-P)^7 + (3f_n+25)P^2(1-P)^6 + (18f_n+38)P^3(1-P)^5 +$ |
| | $(41f_n+29)P^4(1-P)^4 + (44f_n+12)P^5(1-P)^3 + (\frac{205f_n+47}{9})P^6(1-P)^2 +$ |
| | $(\frac{50f_n+22}{9})P^7(1-P) + \frac{13}{27}P^8$ |
| $C_5$ | $(1-P)^8 + 8P(1-P)^7 + 26P^2(1-P)^6 + 44P^3(1-P)^5 + \frac{128}{3}P^4(1-P)^4$ |
| | $\frac{80}{3}P^5(1-P)^3 + \frac{346}{27}P^6(1-P)^2 + \frac{116}{27}P^7(1-P) + \frac{13}{27}P^8$ |
| $C_7$ | $(1-P)^8 + 8P(1-P)^7 + \frac{728}{27}P^2(1-P)^6 + \frac{448}{9}P^3(1-P)^5 + \frac{4522}{81}P^4(1-P)^4$ |
| | $\frac{3304}{81}P^5(1-P)^3 + \frac{14672}{729}P^6(1-P)^2 + \frac{4432}{729}P^7(1-P) + \frac{13}{27}P^8$ |
| $C_9$ | $(1-P)^8 + 8P(1-P)^7 + \frac{55}{2}P^2(1-P)^6 + 53P^3(1-P)^5 + \frac{379}{6}P^4(1-P)^4$ |
| | $\frac{146}{3}P^5(1-P)^3 + \frac{1307}{54}P^6(1-P)^2 + \frac{191}{27}P^7(1-P) + \frac{13}{27}P^8$ |

**Table** 8 : Fidelity without and with correction by the five, seven and nine qubits codes. The symbol C$_0$ corresponds to no correction.

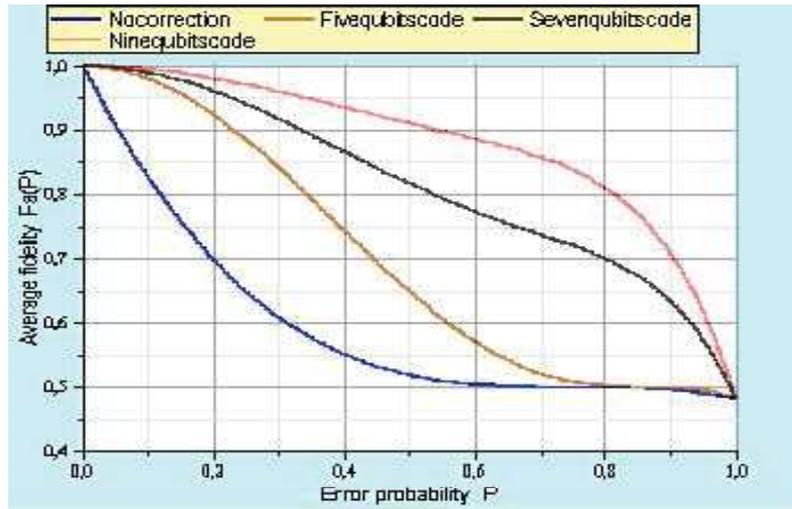

**Figure** 4 : Fidelity without and with correction by the five, seven and nine qubits codes.

## 9 Conclusion

This work was devoted to error correction for quantum secret sharing. The results show that the nine qubits code gives the best fidelity, followed by the seven, then the five qubits code, regardless the depolarizing channel error probability $P$. This conclusion seems to confirm the simulation work done in [11], where errors were introduced by the correction process itself. We conclude that higher is the ancillas number better is the fidelity. The reason is that the nine qubits code offers a higher fraction of double channel errors letting unaffected the received useful qubit. In fact, as only single and double errors have been considered, the nine qubits code gives a specific syndrome for a higher number of double errors, then allowing their recovery which lead to fidelity equal to one. We have supposed that triple errors and more are very unlikely and then with negligible effect on the obtained results.

## Acknowledgement

We would like to thank very much Mrs S.Fritzsch and T.Radtke for providing us with the version 4 (2008) of Feynman Program.

## 11 References

[1] Graph States for Quantum Secret Sharing, Damian Markham and Barry C. Sanders, Phys. Rev. A 78, 042309 (2008).
[2] T.Radtke, S.Fritzsche: 'Simulation of n-qubits quantum systems', I. Quantum gates and registers, CPC, Volume 173, Issues 1–2, 2005, Pages 91–113.
[3] T.Radtke, S.Fritzsche: 'Simulation of n-qubits quantum systems', II. Separability and entanglement, CPC, Volume 175, Issue 2, 2006, Pages 145–166.
[4] T.Radtke, S.Fritzsche: 'Simulation of n-qubits quantum systems', III. Quantum operations, CPC, Volume 176, Issues 9–10, 2007, Pages 617–633.
[5] T.Radtke, S.Fritzsche: 'Simulation of n-qubits quantum systems',IV. Parametrizations of quantum states, CPC, Volume 179, Issue 9, 2008, Pages 647–664.
[6] CPC Prog Lib, Queen's University of Belfast, N.Ireland, Apr 2008.
[7] M.A. Nielsen, I.L Chuang: "Quantum computation and information", Cambridge University Press, UK, 2000.
[8] Raymond Laflamme and co "Perfect Quantum Error Correcting Code", Physic Review Letters, Volume 77, Number 1,198-201, july 1996, .
[9] A. M. Steane. "Multiple particle interference...". Proc. R. Soc. Lond. A, 452:2551–2576, 1996. quant-ph/9601029.
[10] Austin G. Fowler, "Constructing arbitrary Steane code single logical qubit fault-tolerant gates", Quantum Information and Computation,11: 867-873 (2011).
[11] Jumpei Niwa, Keiji Matsumoto, Hiroshi Imai, "Simulating the Effects of Quantum Error-correction Schemes", arXiv:quant-ph/0211071v1 13 Nov 2002.